\documentclass[%
 reprint,
 amsmath,amssymb,
 aps,
]{revtex4-2}
\UseRawInputEncoding
\usepackage{graphicx}
\usepackage{dcolumn}
\usepackage{enumerate}
\usepackage{amsmath}
\usepackage{makecell}
\usepackage{adjustbox}
\usepackage{float}
\usepackage{bm}

\begin{document}

\preprint{APS/123-QED}

\title{Phase space analysis of R\'enyi Holographic dark energy model }

\author{
    Santanu Das\textsuperscript{1} \\
    \textsuperscript{1}Department of Basic Science and Humanities, Institute of Engineering \& Management, University of Engineering and Management, Kolkata, India.\\
    Department of Mathematics, Jadavpur University, Kolkata, India\\ Email: santanudas.ju@gmail.com
    \\and
    Nilanjana Mahata\textsuperscript{2} \\
    \textsuperscript{2}Department of Mathematics, Jadavpur University, Kolkata, India.\\ Email: nilanjana.mahata@jadavpuruniversity.in
}



\begin{abstract}
 Recent observational evidences point out towards a late time acceleration of the universe. In order to study the accelerated expansion, scientists have incorporated the existence of an exotic matter with negative pressure, termed as  dark energy. Afterwards a new idea of dark energy has been studied depending on the holographic principle of quantum gravity, called as the Holographic Dark Energy(HDE). Later on modifying Bekestein-Hawking entropy, different generalized entropies have been proposed, one of them being R\'enyi entropy which leads to R\'enyi holographic dark energy model (RHDE). We have considered RHDE model with Hubble horizon as the IR cut off and have studied  the cosmological behaviour under non interacting, linear and non-linear interacting scenarios with the help of dynamical systems analysis. We have also investigated the stability of the system around hyperbolic critical points along with the type of fluid description, evolution of equation of state parameter as well as matter and energy density parameters.
\end{abstract}

\maketitle


\section{\label{sec:level I}Introduction:\protect}

Observational evidences over the last few decades  such as type I supernova \cite{nh,tonry,riess2007,me,se,id,co,Aldering2002SNAP}, large scale structure (LSS) \cite{percival2007bao,tegmark2004cosmological,t3,de}, CMBR anisotropes \cite{fi,fy,ty,WMAP:2006bqn,spergel2003wmap,goldstein2003acbar,sy,mt,tc,Corasaniti2004} etc. indicate  that our universe is  having  accelerated expansion at present. There are two distinct approaches   to incorporate  this  late time accelerated expansion. One approach  is   to modify  Einstein gravity  in terms of geometrical effects  resulting  various models of Modified theories of gravities such as f(R), f(R,T), f(Q) etc  \cite{cha,sn,bo,bra,ens,amus,nm,ss,singh2014frt}. Another approach  is  to modify the right hand side  of Einstein Field equation i.e modifying the matter part  yielding different  dark energy models. In this later  case, an unknown form of energy having significant negative pressure called dark energy (DE) is introduced in the Einstein equation which can achieve a negative equation of state parameter \cite{td,ni,gj,iz,cf,padmanabhan2005dark,copeland2006dynamics,amendola2010dark,sahni2004dark,mahata2014phantom}. Though Cosmological constant is the simplest and most suitable candidate for dark energy, having constant equation of state $\omega = -1$, it suffers from two major problems, namely, coincidence problem and fine tuning problem. To counter these problems, scintists proposed various dynamical dark energy models such as quintessence, k-essence, other scalar field models, holographic dark energy (HDE) models etc. Further, interacting dark energy models have drawn much interest as it has been shown that proper choice of interaction term between dark matter and dark energy may alleviate the coincidence problem \cite{Cohen1999,Li2004,Horvat2004,Mahata2015,Hsu2004,Guberina2007,Xu2009,Pavon2005,li2013holographic,delcampo2011holographic,pourojaghi2021cosmography,li2015grde}.\\
Several  dynamical dark enery models have been suggested  to explain the current scenario of the universe. Among which, holographic dark energy models play an important role in explaining the riddle of dark energy. Holographic dark energy models which is based on holographic principle, have received considerable attentions from researchers in context of quantum gravity. \\
 In a cosmological context, the holographic principle imposes a fundamental upper limit on the total entropy. According to holographic principle, the total entropy enclosed within a region of size L must not surpass that of a black hole of the same scale. This implies that the vacuum energy within such a region must be constrained to avoid gravitational collapse, thereby maintaining consistency with the holographic bound. Guided by the Bekenstein bound, it is natural to impose that, for an effective quantum field theory confined to a region of size L with an ultraviolet (UV) cutoff $\Lambda$, the total entropy should respect the holographic limit — meaning it must not exceed the entropy of a black hole occupying the same volume. The total entropy should satisfy the relation
  \begin{equation}
      L^{3}\Lambda^{3}\leq S_{BH}=\pi L^{2} M_{p}^{2}
  \end{equation}
  where $M_{p}$ is the reduced plank mass and $S_{BH}$ is the entropy of a black hole with radius L which is assumed as a long distance IR cut off.  Miao Li \cite{Li2004} had suggested a more rigorous bound which tells that total energy of a region of size L must not surpass the mass of a black hole with same size.\\
  As a result this UV-IR relation provides an upper bound on the zero point energy density as
\begin{equation}
     \rho_{\Lambda} \leq L^{-2}M_{p}^{2}
  \end{equation}
  Thus the holographic dark energy density is obtained as 
  \begin{equation}
     \rho_{\Lambda} =\frac{3c^{2}M_{p}^{2}}{L^2}
  \end{equation}
where c is a free dimensionless parameter.\\
It seems there is deep connection between dark energy, horizon entropy and  laws of thermodynamics. 
Thus horizon entropy and dark energy candidates may influence one another from a thermodynamic perspective. Recently, generalized entropy formalisms have been widely used to study various cosmological and gravitational scenarios because of the unknown nature of spacetime, the long-range nature of gravity and the fact that the Bekenstein–Hawking entropy is a non-extensive entropy measure.\\
Amongst them the R\'enyi \cite{Renyi1961}, Tsallis \cite{Tsallis2013}, Sharma– Mittal \cite{sharma1975,sharma1977}, kanidakis \cite{Kaniadakis2002} entropies acquired considerable interest. 
Recently, using the R\'enyi entropy \cite{Renyi1961}, Moradpour et al. has suggested a new type of  DE model, called the R´enyi holographic dark energy (RHDE) model \cite{Moradpour2018}, to understand the  gravitational and cosmological incidences. Moradpour et al. has shown that if there is no interaction between the dark matter and energy, RHDE model is more stable. 
Also, there are various other works on the RHDE \cite{ maity2019, jawad2018,dubey2020,prasanthi2020,VijayaSanthi2022}. 

The effectiveness of these attempts to describe the current accelerating universe encourages us to study cosmic evolution considering    generalized entropies as the horizon entropy instead of the Bekenstein entropy.  We will use this formalism to consider a new model for HDE in flat FRW by taking the Hubble radius as its IR cutoff with R\'enyi generalized entropy, since it has been demonstrated that generalized entropy formulation can produce a suitable description for dark energy.\\
In this paper, we have discussed  basic R\'enyi entropy and  R\'enyi holographic dark energy in section-II, basic Friedmann equations pertaining to R\'enyi entropy and R\'enyi HDE in section-III. We have discussed R\'enyi HDE model without interaction along with the stability analysis of non hyperbolic critical point in section-IV,  R\'enyi HDE model with interaction where we have examined three linear interactions  $Q=3H\rho$, $Q=3H\rho_d$, $Q=3H\rho_m$ and a non linear interaction   $Q=3H\frac{\rho_d}{\rho_d+\rho_m}$ in section-V. Section-VI contains comparison with other RHDE models and section-VII has the conclusion.

\section{\label{sec:level II}Introduction of R\'enyi entropy in  Holographic dark energy:\protect}

R\'enyi holographic dark energy model \cite{Manoharan2023,Moradpour2018} is connected to R\'enyi entropy\cite{Renyi1970,Renyi1961}. For a system of $w$ discrete states, it is defined by  
\begin{equation}
S_R= k\frac{ln\sum_{i=1}^w p_{i}^q}{1-q} 
\end{equation}
where $q$ is a non extensive parameter, $k$ is a positive constant and $p_i$ is the probability for $i$th state. R\'enyi entropy reduces to standard Boltzmann- Gibbs entropy for $\lim q \to 1$.
Using Tsallis entropy defined as 
\begin{equation}
S_T=k \frac{1-\sum_{i=1}^w p_{i}^q}{q-1}
\end{equation}

Relation between  Tsallis entropy and R\'enyi entropy can be written as 
\begin{equation}
S_R=\frac{1}{\delta}ln(1+\delta S_T) 
\end{equation}
where $\delta=1-q$ is a parameter which measures whether the system is non-additive  or not.
By deducing (6) from \cite{Biro2011}, it was shown that when formal logarithmic transformations are applied to generalized thermodynamic quantities, the resulting expressions exhibit additive behavior. These transformed forms continue to follow the conventional laws and relationships of standard thermodynamics. To calculate the R\'enyi entropy related to black hole \cite{Czinner2016Renyi,Biro2013,Komatsu2017,Moradpour2018Generalized}, Tsallis entropy is considered  as formal Bekenstein-Hawking entropy. As a result (6) is redefined as 
\begin{equation}
S_R=\frac{1}{\delta}ln(1+\delta S_{BH}) 
\end{equation}
where $S_{BH}$ is Bekenstein-Hawking entropy which is denoted as 
\begin{equation}
S_{BH}=\frac{k_Bc^3A}{4 \hbar G} 
\end{equation}
where $k_B$, $c$, $A$, $\hbar$, $G$ are respectively Boltzmann constant, speed of light, area of the black hole horizon, reduced Plank constant and Newton's gravitational constant. By considering 
in natural units where $ k_B = \hbar = c = G = 1 $, it simplifies to:
\begin{equation}
S_{BH} = \frac{A}{4} 
\end{equation}
By assuming $A=4\pi L^2$, where $L$ is the IR cut off and using (9), equation (7) is rewritten as 
\begin{equation}
S_R=\frac{1}{\delta}ln(1+\pi \delta L^2 )
\end{equation}
Now, using the thermodynamic relation 
\begin{equation}
T dS_R \propto  \rho_d dV 
\end{equation}
R\'enyi holographic dark energy density is defined as
\begin{equation}
\rho_d = \frac{3 d^2}{8 \pi  L^2} (1+\pi \delta L^2)^{-1} 
\end{equation}
where $d$ is a constant, $\delta$ is the non-extensive R\'enyi parameter, and $L$ is the IR cutoff.\\
  In this work, we will consider R\'enyi holographic dark energy with Hubble's cut off.\\
   So, we will assume 
  \begin{equation}
  L=\frac{1}{H}
  \end{equation}
  where $H$ is the Hubble's parameter.
  Considering equation (13), equation (12) is transformed into  
 \begin{equation}
  \rho_{d} =\frac{3d^2}{8\pi}\frac{H^4}{H^2+\pi\delta}
 \end{equation}

 \section{\label{sec:level III}Basic equations:\protect}
 We assume the universe is homogeneous and isotropic on large scales and is described by the spatially flat Friedmann-–Lemaître-–Robertson-–Walker (FLRW) metric with the line element:
 \begin{equation}
     ds^{2}=dt^2-a^2(t)(dr^2+r^2d\Omega^2)
 \end{equation}
 where $a(t)$ is the scale factor and $d\Omega^2=d\theta^2+sin^2 \theta d\phi^2$ which denotes a two dimensional sphere.\\
 In FLRW cosmology matter density will follow the energy-momentum tensor pertaining to the perfect fluid, which is denoted by 
  \begin{equation}
    T_{\mu\nu}=(p+\rho)u_{\mu}u_{\nu}-p g_{\mu\nu}
 \end{equation}
 In this framework, $u_{\mu}$ is the four-velocity of the fluid, satisfying the normalization condition satisfying $u^{\mu}u_{\mu}=1$ while the cosmic objects are characterized by their energy density $\rho$ and pressure $p$. Here, we assume that the universe is filled with dark matter in the form of dust and dark energy in the form of R\'enyi hologrpahic dark energy with variable equation of state.\\ 
 The Einstein field equations pertaining to the current cosmological model can be written as 
 \begin{equation}
 3H^2=\rho=\rho_{m}+\rho_{d}
 \end{equation}
  \begin{equation}
 2 \dot{H}+3H^2=-p=-p_{m}-p_{d}
 \end{equation}
 Equation (18) can be rewritten as 
   \begin{equation}
 2\dot{H}=-\rho_{m}-(1+\omega_{d})\rho_{d}
 \end{equation}
 Here $\rho_{m}$, $\rho_{d}$ are the energy density of dark matter and holographic dark energy density, $p_{m}$, $p_{d}$ denote the pressure of dark matter and holographic dark energy respectively. $\omega_{d}$ is the variable equation of state parameter pertaining to the holographic dark energy which is defined as $\omega_{d}=\frac{p_{d}}{\rho_{d}}$.\\
 Using (17) and (19), acceleration of the universe is denoted by 
  \begin{equation}
 \ddot{a}=-\frac{a}{6}[\rho_{m}+p_{d}(1+3\omega_{d})]
 \end{equation}
 which shows that for cosmic acceleration, $\omega_{d}<-\frac{1}{3}$.
 Dark matter and dark energy components satisfy the continuity equation individually as 
 \begin{equation}
 \dot{\rho_{m}}+3H\rho_{m}=0
 \end{equation}
 and 
 \begin{equation}
 \dot{\rho_{d}}+3H(1+\omega_{d})\rho_{d}=0
 \end{equation}
  Considering the interaction term between two dark components, the continuity equations transform into 
  \begin{equation}
 \dot{\rho_{m}}+3H\rho_{m}=Q
 \end{equation}
  \begin{equation}
 \dot{\rho_{d}}+3H(1+\omega_{d})\rho_{d}=-Q
 \end{equation}
 The interaction term $Q$ is not uniquely determined; however, we adopt the choice $Q>0$, implying an energy flow from dark energy to dark matter i.e., dark energy decays into dark matter. This positive sign of $Q$ supports the second law of thermodynamics by ensuring non-decreasing entropy and also contributes toward addressing the coincidence problem. It is worth mentioning that baryonic matter is excluded from the interaction, as such couplings are tightly constrained by local gravitational observations \cite{Jain2010,Will2014,Clifton2012}.\\
 In our current work, we have considered three different types of interaction terms  which include linear and non-linear interaction terms, namely:
 \begin{itemize}
  \item $Q=3H(\rho_{m}+\rho_{d}$)
  \item $Q=3H\rho_{d}$.
  \item $Q=3H\rho_{m}$.
  \item $Q=3H\frac{\rho_{d}}{\rho_{m}+\rho_{d}}$.
\end{itemize}

 \section{\label{sec:level IV}R\'enyi Holographic dark energy model with Hubble horizon as IR cut off (without interaction) :\protect}
In the present study, we explore a dynamical dark energy comprising both dark energy and dark matter, under the assumption that there is no interaction between them. The dark energy component is described by the R\'enyi holographic dark energy model which is the exclusive source of dark energy where  the Hubble horizon is employed as the infrared (IR) cutoff.  \\
We would like to study aforesaid model by using dynamical system tools as the field equations derived here will be very complex and non-linear in nature.\\
Let us introduce the dimensionless variables :
\begin{equation}
x=\frac{\rho_{m}}{3H^2},~ y=\frac{\rho_{d}}{3H^2}
\end{equation}
Here matter and energy density parameters are defined as 
\begin{equation}
\Omega_{m}=\frac{\rho_{m}}{3H^2}=x,~ \Omega_{d}=\frac{\rho_{d}}{3H^2}=y
\end{equation}
Using equation(17) and (26) we can deduce that 
\begin{equation}
\Omega_{m}+\Omega_{d}=x+y=1
\end{equation}
By employing equations (17), (21), (22), (25) and (27) we derive the corresponding autonomous system of differential equations as follows:
\begin{equation}
\begin{split}
\frac{dx}{dN}=3xy\omega_{d} \\
\frac{dy}{dN}=-3y(1-y)\omega_{d}
\end{split}
\end{equation}
Using (14), from (22), we can obtain equation of state parameter $\omega_{d}$ for R\'enyi HDE as 
\begin{equation}
\omega_{d}=-1-\frac{4}{3}\frac{\dot{H}}{H^2}+\frac{2}{3}\frac{\dot{H}}{H^2+\pi\delta}
\end{equation}
From (19), we deduce the value of $\frac{\dot{H}}{H^2}$ as
\begin{equation}
\frac{\dot{H}}{H^2}=-\frac{3}{2}(1+y \omega_{d})
\end{equation}
Using (30) in (29), equation of state parameter pertaining to R\'enyi HDE can be simplified as 
\begin{equation}
\omega_{d}=\frac{1+\frac{2}{3}\frac{\dot{H}}{H^2+\pi\delta}}{1-2y}
\end{equation}
For the purpose of analyzing the cosmological model, we have to transform the nonlinear autonomous system presented in equation (28) into a simplified form by incorporating the substitution from equation (31). For such purpose, we are introducing a function of dimensionless variables $x$ and $y$ as 
\begin{equation}
\lambda(x,y)=\frac{\dot{H}}{H^2+\pi\delta}
\end{equation}
In this section, where we study the Rényi HDE model without interaction, we consider two different choices of $\lambda(x,y)$  as:
 \begin{itemize}
  \item $\lambda(x,y)=\alpha x+ \beta y$.
  \item $\lambda(x,y)=e^{\alpha x+ \beta y}$.
\end{itemize}
Here $\alpha$  and $\beta$ are two arbitrary constants.

\subsection{\label{sec:level I} Analysis of non-interacting R\'enyi HDE with $\lambda(x,y)=\alpha x+ \beta y$ :\protect}

Substituting $\omega_d$ from (31) and considering $\lambda(x,y)=\alpha x+ \beta y$, (28) reduces to 
\begin{align}
\frac{dx}{dN}=\frac{3xy+2xy(\alpha x+ \beta y) }{1-2y}\\
\frac{dy}{dN}=\frac{-3y(1-y)-2y(1-y)(\alpha x+ \beta y) }{1-2y}\\
\omega_{d}=\frac{1+\frac{2}{3}(\alpha x+ \beta y)}{1-2y}
\end{align}

To perform dynamical system analysis, we need to identify the critical points corresponding to the autonomous system (33), (34). To find out the critical point of the system we need to solve the system of algebraic equations :
\begin{equation}
\begin{split}
\frac{dx}{dN}=0\\
\frac{dy}{dN}=0
\end{split}
\end{equation}
The critical points of the said autonomous system (33) and (34) have been listed in Table-I.
\begin{table}[htbp]
    \centering
    \scriptsize
    \begin{tabular}{|c|c|}
        \hline
        \textbf{Critical Points} & \textbf{Coordinates $(x, y)$} \\ \hline
        $A$ & $(0, 1)$ \\ \hline
        $A_1$ & $(1, 0)$ \\ \hline
        $A_2$ & $\left(\frac{3\beta}{2\alpha^2}, 0\right)$ \\ \hline
        $A_3$ & $\left(0, -\frac{3}{2\beta}\right)$ \\ \hline
        $A_4$ & $\left(\frac{\beta + \frac{3}{2}}{\beta - \alpha}, \frac{-\alpha - \frac{3}{2}}{\beta - \alpha}\right)$ \\ \hline
    \end{tabular}
    \caption{Set of critical points and their coordinates.}
\end{table}
The eigenvalues corresponding to each critical point are obtained from the linearized Jacobian matrix and are summarized in Table II.
\begin{table}[htbp]
    \centering
    \scriptsize
    \begin{tabular}{|c|c|}
        \hline
        \textbf{Critical Points} & \textbf{Eigenvalues} \\ \hline
        $A$     & $\left(-3 - 2\beta, -3 - 2\beta\right)$ \\ \hline
        $A_1$   & $\left(-3 - 2\alpha, 0\right)$ \\ \hline
        $A_2$   & $\left(0, \frac{-3\alpha - 3\beta}{\alpha}\right)$ \\ \hline
        $A_3$   & $\left(0, \frac{9 + 2\beta}{6 + 2\beta}\right)$ \\ \hline
        $A_4$   & $\left(0, \frac{6\alpha + 6\beta + 4\alpha\beta}{2\beta + 2\alpha + 6}\right)$ \\ \hline
    \end{tabular}
    \caption{Critical points and their corresponding eigenvalues from the linearized Jacobian matrix.}
\end{table}
Let us now study the characteristics of the critical points along with their stability criteria. We will also study the behavior of physical parameters corresponding to the critical points.
\begin{itemize}
  \item Critical point $A$ becomes a stable node when $\beta>-\frac{3}{2}$ and unstable otherwise. Equation of the state parameter $\omega_{d}<0$ when $A$ is stable. For $\beta=0$, we get $\omega_{d}=-1$ i.e., the stable node represents $\Lambda$CDM. The fluid description is like phantom pertaining to R\'enyi HDE when $\beta>0$ and the fluid description is like quintessence when $-\frac{3}{2}<\beta<0$. The universe experiences an accelerated expansion when $\beta>-1$. So, the stable node always represents with phantom or quintessence like fluid description of universe pertaining to R\'enyi HDE without interaction.\\
  Around the critical point, universe is completely dark energy dominated as the matter density, $\Omega_{m}=0$ and the dark energy density pertaining to HDE, $\Omega_{d}=1$.
	\begin{figure}[h!]
\centerline{\includegraphics[scale=0.6]{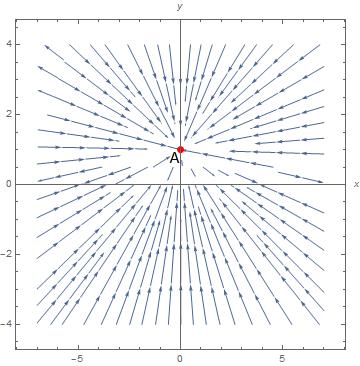}}
\caption{Phase plot corresponding to the point $(0,1)$ for $\alpha=0.2, \beta=-0.1$}
\end{figure}
  Fig-1 shows that for $\beta=-0.1$, $A$ is a stable node with equation of state parameter $\omega_{d}=-0.93$ which depicts the quintessence like fluid with accelerated expansion. The parameter $d$ which is there in the R\'enyi holographic dark energy density corresponding to (14) does not impact much in our model. But $\frac{\pi\delta}{H^2}$ can be evaluated by using (31) and (32) as $-0.3$ at the critical point $A$ for $\beta=-0.1$ and $w_{d}=-0.93$.
\begin{figure}[h!]
\centerline{\includegraphics[scale=0.6]{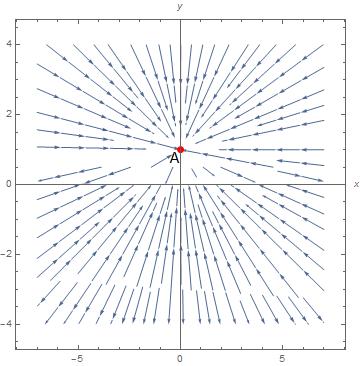}}
\caption{Phase plot corresponding to the point $(0,1)$ for $\alpha=0.02, \beta=0.5$}
\end{figure}
  Fig-2 shows that for $\beta=0.5>0$, $A$ represents a stable node with equation of state parameter $\omega_{d}=-1.33$ which depicts the phantom like fluid with accelerated expansion. Here also $\frac{\pi\delta}{H^2}$ can be evaluated by using (31) and (32) as $-0.34$ at the critical point $A$ for $\beta=0.5$ and $w_{d}=-1.33$. So, we can conclude here that the value of $\delta$ is negative.

\begin{figure}[h!]
\centerline{\includegraphics[scale=0.6]{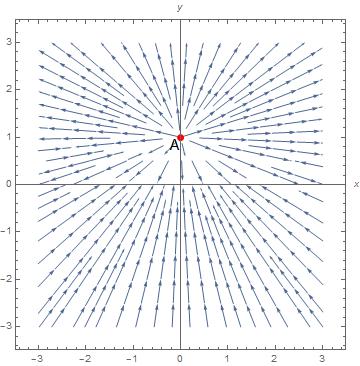}}
\caption{Phase plot corresponding to the point $(0,1)$ for $\alpha=-0.2, \beta=-2.5$}
\end{figure}
  Fig-3 shows that for $\beta=-2.5<-\frac{3}{2}$, $A$ represents an unstable node with equation of state parameter $\omega_{d}=0.66$ .
\begin{figure}[h!]
\centerline{\includegraphics[scale=0.6]{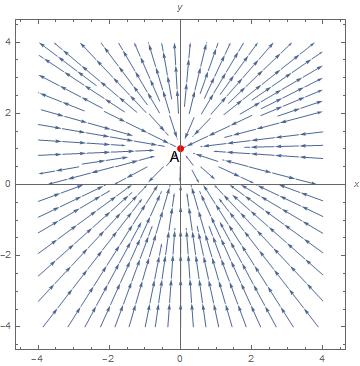}}
\caption{Phase plot corresponding to the point $(0,1)$ for $\alpha=-0.02, \beta=-0.75$}
\end{figure}
  Fig-4 shows that for $\beta=-0.75$, $A$ represents a stable node with equation of state parameter $\omega_{d}=-0.5$ which depicts the quintessence like fluid with accelerated expansion. \\
  So, more we are shifting the value $\beta$ from $0$ towards $-1$, the value of equation of state parameter increases from $-1$ towards $-0.33$ and represents the quintessence like fluid description with accelerating nature of the universe.\\
  
  \begin{figure}[h!]
\centerline{\includegraphics[scale=0.6]{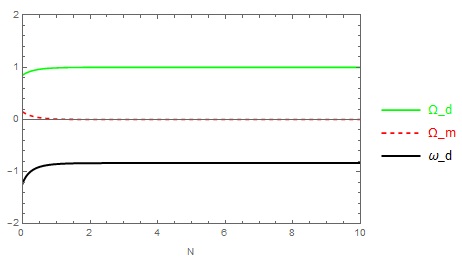}}
\caption{Evolution of cosmological parameters corresponding to the point $(0,1)$ for $\alpha=0.2, \beta=-0.25$}
\end{figure}
Fig-5 shows that there is a transition from phantom to quintessence era and it also shows the existence of dark energy dominated era and the source of the dark energy is influenced here by R\'enyi HDE.
\item Critical points $A_{1},~A_{2},~A_{3},~A_{4}$ are non hyperbolic in nature as they are having one of the eigen value with zero real part. So we can not analyze them because the current linearization tools which we are using here fail to provide any conclusive information regarding non-hyperbolic points. 
\end{itemize}

\subsection{\label{sec:level II} Analysis of non-interacting R\'enyi HDE with $\lambda(x,y)=e^{\alpha x+ \beta y}$ :\protect}

Substituting $\omega_d$ from (31) and considering $\lambda(x,y)=e^{\alpha x+ \beta y}$, (28) reduces to 
\begin{align}
\frac{dx}{dN}=\frac{3xy+2xye^{\alpha x+ \beta y} }{1-2y}\\
\frac{dy}{dN}=\frac{-3y(1-y)-2y(1-y)e^{\alpha x+ \beta y} }{1-2y}\\
\omega_{d}=\frac{1+\frac{2}{3}e^{\alpha x+ \beta y}}{1-2y}
\end{align}

To perform dynamical system analysis, we need to identify the critical points corresponding to the autonomous system (37), (38). To find out the critical point of the system we need to solve the system of algebraic equations :
\begin{equation}
\begin{split}
\frac{dx}{dN}=0\\
\frac{dy}{dN}=0
\end{split}
\end{equation}
The critical points of the said autonomous system (37) and (38) have been listed in Table-III.

\begin{table}[htbp]
    \centering
    \scriptsize
    \begin{tabular}{|c|c|}
        \hline
        \textbf{Critical Points} & \textbf{Coordinates $(x, y)$} \\ \hline
        $B$     & $(0, 1)$ \\ \hline
        $B_1$   & $(1, 0)$ \\ \hline
    \end{tabular}
    \caption{Set of critical points and their coordinates.}
\end{table}

The eigenvalues corresponding to each critical point are obtained from the linearized Jacobian matrix and are summarized in Table IV.
\begin{table}[htbp]
    \centering
    \scriptsize
    \begin{tabular}{|c|c|}
        \hline
        \textbf{Critical Points} & \textbf{Eigenvalues} \\ \hline
        $B$   & $\left(-3 - e^{2\beta},\ -3 - e^{2\beta}\right)$ \\ \hline
        $B_1$ & $\left(0,\ -3 - e^{2\alpha}\right)$ \\ \hline
    \end{tabular}
    \caption{Critical points and their corresponding eigenvalues from the linearized Jacobian matrix.}
\end{table}

Now, we investigate the characteristics of the critical points along with their stability criteria. We will also study the behavior of physical parameters corresponding to the critical points.

\begin{itemize}
  \item Critical point $B$ becomes a stable node for any $\beta \in \mathbb{R}$. Equation of the state parameter is always  $\omega_{d}<-1$. So, the fluid description is always like phantom pertaining to R\'enyi HDE. System around this critical point always indicates towards the accelerated expansion of the universe.\\
  Around the critical point, universe is completely dark energy dominated as the matter density, $\Omega_{m}=0$ and the dark energy density pertaining to HDE, $\Omega_{d}=1$.
	\begin{figure}[h!]
\centerline{\includegraphics[scale=0.6]{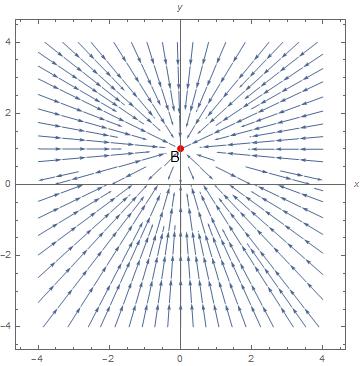}}
\caption{Phase plot corresponding to the point $(0,1)$ for $\alpha=2, \beta=1$}
\end{figure}
  Fig-6 shows that  $B$ is a stable node with equation of state parameter $\omega_{d}=-2.81$ which depicts the phantom like fluid with accelerated expansion.
\begin{figure}[h!]
\centerline{\includegraphics[scale=0.6]{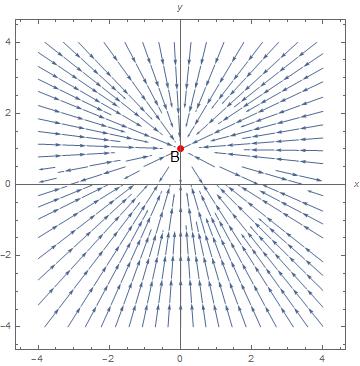}}
\caption{Phase plot corresponding to the point $(0,1)$ for $\alpha=0.25, \beta=-0.75$}
\end{figure}
  Fig-7 shows that $B$ represents a stable node with equation of state parameter $\omega_{d}=-1.31$ which depicts the phantom like fluid with accelerated expansion.\\
  More over, we find that as the value of the parameter $\beta$ decreases from 1, equation of state parameter $\omega_{d}$ increases towards -1.

   \begin{figure}[h!]
\centerline{\includegraphics[scale=0.6]{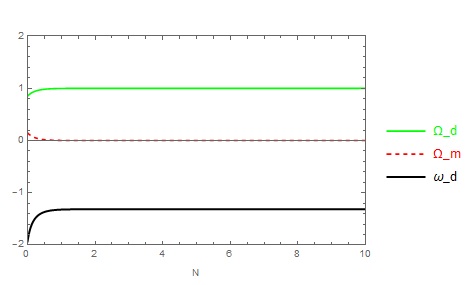}}
\caption{Evolution of cosmological parameters corresponding to the point $(0,1)$ for $\alpha=0.25, \beta=-0.75$}
\end{figure}
Fig-8 shows that equation of state parameter increases towards -1 and it also shows the existence of dark energy dominated era and the source of the dark energy is influenced here by R\'enyi HDE.
\item Critical point $B_{1}$ is  non hyperbolic in nature as it possesses a eigen value with vanishing real part. So we are not studying it's nature in details in our present work.
\end{itemize}

    

\section{\label{sec:level V}R\'enyi Holographic dark energy model with Hubble horizon as IR cut off (with interaction) :\protect}

In this section, we would like to consider interaction between matter and energy and we want to study the behavior of our cosmological model. Here, we are considering both linear and nonlinear interaction without coupling constant.

\subsection{\label{sec:level II} Analysis of interacting R\'enyi HDE with interaction $Q=3H(\rho_m+\rho_d)$ :\protect}

Here, we study our said model with R\'enyi HDE upon considering a linear interaction $Q$ in the form of 
\begin{equation}
Q=3H(\rho_{m}+\rho_{d}).
\end{equation}
By using (41), equation (23),(24) transformed into 
\begin{equation}
 \dot{\rho_{m}}=3H\rho_{d}
 \end{equation}
 and 
 \begin{equation}
 \dot{\rho_{d}}=-6H\rho_{d}-3H\rho_{d}\omega_{d}-3H\rho_{m}
 \end{equation}
By incorporating the dimensionless variables defined in (25) and using the equations (30),(42) and (43), we obtain the following system of autonomous equations  

\begin{equation}
\begin{split}
\frac{dx}{dN}=3+3xy\omega_{d} \\
\frac{dy}{dN}=-3-3y(1-y)\omega_{d}
\end{split}
\end{equation}
Using (14) and (30), from (43), we obtain the equation of state parameter $\omega_{d}$ for R\'enyi HDE as 
\begin{equation}
\omega_{d}=\frac{\frac{2}{3}\frac{\dot{H}}{H^2+\pi\delta}-\frac{x}{y}}{1-2y}
\end{equation}
Here, we assume as earlier $\lambda(x,y)=\frac{\dot{H}}{H^2+\pi\delta}$.
Here, we are assuming only one choice of 
\begin{equation}
\lambda(x,y)=\alpha x+\beta y
\end{equation}
as choice of an exponential function is not giving us any viable critical points.
By using (45) and  (46), our system of autonomous equations (44) transforms into 
\begin{equation}
\begin{split}
\frac{dx}{dN}=3+\frac{2xy(\alpha x+\beta y)-3x^2}{1-2y} \\
\frac{dy}{dN}=-3+\frac{-2y(1-y)(\alpha x+\beta y)+3x(1-y)}{1-2y} 
\end{split}
\end{equation}

The critical points of this  autonomous system  are enlisted in Table-V.

\begin{table}[htbp]
    \centering
    \scriptsize
    \resizebox{\linewidth}{!}{
    \begin{tabular}{|c|c|}
        \hline
       \tiny{\textbf{Critical}} & \tiny{\textbf{Coordinates}}\\
        \tiny{\textbf{Points}} & \ \tiny{\textbf{$(x, y)$}} \\ \hline
        \tiny{$C$} &  \tiny{$(1, 0)$} \\ \hline
         \tiny{$D_{1}$} &  \tiny{$(1-\frac{3+4\alpha-2\beta+\sqrt{9+24\alpha-12\beta+4\beta^2}}{4(\alpha-\beta)}, \frac{3+4\alpha-2\beta+\sqrt{9+24\alpha-12\beta+4\beta^2}}{4(\alpha-\beta)})$} \\ \hline
        \tiny{$D_{2}$} &  \tiny{$(1-\frac{3+4\alpha-2\beta-\sqrt{9+24\alpha-12\beta+4\beta^2}}{4(\alpha-\beta)}, \frac{3+4\alpha-2\beta-\sqrt{9+24\alpha-12\beta+4\beta^2}}{4(\alpha-\beta)})$} \\ \hline
     
    \end{tabular}
    }
    \caption{Set of critical points and their coordinates.}
\end{table}

Now, we study the characteristics of the system around the critical points and we will also study the evolution of the cosmological parameters.

\begin{itemize}
  \item  For critical point $C$, we evaluate the eigen values from the linearized Jacobian matrix which is formed from the autonomous system (47). Both the eigen values corresponding to $C$  are $(-3,-2\alpha)$ which indicates that $C$ is a stable node for any $\alpha>0$. Equation of the state parameter is undefined here. So, we can't conclude anything regarding acceleration corresponding to the said critical point.\\
  Around the critical point, universe is completely dark matter dominated as the matter density, $\Omega_{m}=1$ and the dark energy density pertaining to HDE, $\Omega_{d}=0$.\\
	\begin{figure}[h!]
\centerline{\includegraphics[scale=0.6]{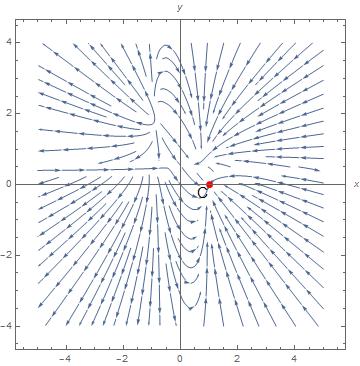}}
\caption{Phase plot corresponding to the point $(1,0)$ for $\alpha=4, \beta=0.25$}
\end{figure}

  Fig-9 shows that  $C$ is locally stable  for $\alpha=4>0$.

\item It is very difficult to calculate the eigen values corresponding to  $D_{1}$ and $D_{2}$. For this we are having different choices of the values of parameters $\alpha,~\beta$ and corresponding to those values, we are representing respective points associated to $D_{1}$ and $D_{2}$ along with their eigen values and different cosmological parametric values in Table-VI and Table-VII respectively.

\begin{table}[htbp]
    \centering
    \scriptsize
    \begin{adjustbox}{width=\linewidth}
    \begin{tabular}{|c|c|c|c|c|c|}
        \hline 
        \tiny \textbf{Choices of $(\alpha,\beta)$} & 
        \tiny \textbf{$D_1$} & 
        \tiny \textbf{Eigen Values} & 
        \tiny \textbf{$\Omega_{m}$} & 
        \tiny \textbf{$\Omega_{d}$} &  
        \tiny \textbf{$\omega_{d}$} \\ \hline 

          \tiny \textbf{$(0,-3)$} & 
        \tiny \textbf{$(-\frac{1}{2},\frac{3}{2})$} & 
        \tiny \textbf{$(6,\frac{27}{4})$} & 
        \tiny \textbf{$-\frac{1}{2}$} & 
        \tiny \textbf{$\frac{3}{2}$} &  
        \tiny \textbf{$1.33$} \\ \hline
        
         \tiny \textbf{$(-4,-3.5)$} & 
        \tiny \textbf{$(-1,2)$} & 
        \tiny \textbf{$(\frac{4}{3},3)$} & 
        \tiny \textbf{$-1$} & 
        \tiny \textbf{$2$} &  
        \tiny \textbf{$0.5$} \\ \hline

          \tiny \textbf{$(\frac{2}{3},3)$} & 
        \tiny \textbf{$(\frac{3}{2},-\frac{1}{2})$} & 
        \tiny \textbf{$(-2,\frac{5}{4})$} & 
        \tiny \textbf{$\frac{3}{2}$} & 
        \tiny \textbf{$-\frac{1}{2}$} &  
        \tiny \textbf{$1.33$} \\ \hline

          \tiny \textbf{$(5,2)$} & 
        \tiny \textbf{$(-\frac{3}{2},\frac{5}{2})$} & 
        \tiny \textbf{$(2,\frac{55}{8})$} & 
        \tiny \textbf{$-\frac{3}{2}$} & 
        \tiny \textbf{$\frac{5}{2}$} &  
        \tiny \textbf{$0.26$} \\ \hline

    \end{tabular}
   \end{adjustbox}
    \caption{Eigen values and value of other cosmological parameters corresponding to $D_{1}$ for different choices of $\alpha$ and $\beta$}
\end{table}

\begin{table}[H]
    \centering
    \scriptsize
    \begin{adjustbox}{width=\linewidth}
    \begin{tabular}{|c|c|c|c|c|c|}
        \hline 
        \tiny \textbf{Choices of $(\alpha,\beta)$} & 
        \tiny \textbf{$D_2$} & 
        \tiny \textbf{Eigen Values} & 
        \tiny \textbf{$\Omega_{m}$} & 
        \tiny \textbf{$\Omega_{d}$} &  
        \tiny \textbf{$\omega_{d}$} \\ \hline 

          \tiny \textbf{$(0,-3)$} & 
        \tiny \textbf{$(1,0)$} & 
        \tiny \textbf{$(-3,0)$} & 
        \tiny \textbf{$1$} & 
        \tiny \textbf{$0$} &  
        \tiny \textbf{Undetermined} \\ \hline
        
         \tiny \textbf{$(-4,-3.5)$} & 
        \tiny \textbf{$(-3,4)$} & 
        \tiny \textbf{$(-\frac{8}{7},1)$} & 
        \tiny \textbf{$-3$} & 
        \tiny \textbf{$4$} &  
        \tiny \textbf{$0.08$} \\ \hline

          \tiny \textbf{$(\frac{2}{3},3)$} & 
        \tiny \textbf{$(\frac{3}{7},\frac{4}{7})$} & 
        \tiny \textbf{$(-\frac{228}{7},-\frac{65}{7})$} & 
        \tiny \textbf{$\frac{3}{7}$} & 
        \tiny \textbf{$\frac{4}{7}$} &  
        \tiny \textbf{$-4.08$} \\ \hline

          \tiny \textbf{$(5,2)$} & 
        \tiny \textbf{$(\frac{1}{3},\frac{2}{3})$} & 
        \tiny \textbf{$(-22,-9)$} & 
        \tiny \textbf{$\frac{1}{3}$} & 
        \tiny \textbf{$\frac{2}{3}$} &  
        \tiny \textbf{$-4.5$} \\ \hline

    \end{tabular}
   \end{adjustbox}
    \caption{Eigen values and value of other cosmological parameters corresponding to $D_{2}$ for different choices of $\alpha$ and $\beta$}
\end{table}

\begin{figure}[h!]
\centerline{\includegraphics[scale=0.6]{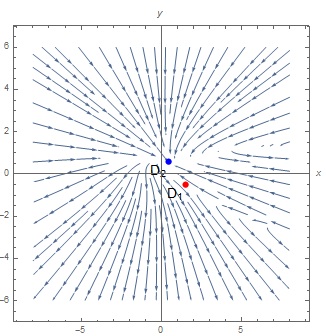}}
\caption{Phase plot corresponding to the point $D_{1}(\frac{3}{2},-\frac{1}{2})$ and  $D_{2}(\frac{3}{7},\frac{4}{7})$ for $\alpha=\frac{2}{3}, \beta=3$}
\end{figure}

  Fig-10 shows that  $D_{1}(\frac{3}{2},-\frac{1}{2})$ is a saddle node  while $D_{2}(\frac{3}{7},\frac{4}{7})$ is locally stable for $\alpha=\frac{2}{3}, \beta=3$.

\begin{figure}[h!]
\centerline{\includegraphics[scale=0.6]{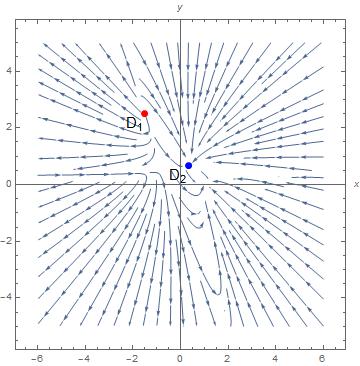}}
\caption{Phase plot corresponding to the point $D_{1}(-\frac{3}{2},\frac{5}{2})$ and  $D_{2}(\frac{1}{3},\frac{2}{3})$ for $\alpha=5, \beta=2$}
\end{figure}
 Fig-11 shows that  $D_{1}(-\frac{3}{2},\frac{5}{2})$ is an unstable node  while $D_{2}(\frac{1}{3},\frac{2}{3})$ is locally stable for $\alpha=5, \beta=2$.\\
  From here we can comment that for positive values of $\alpha$ and $\beta$ , $D_{1}$ is saddle or unstable while $D_{2}$ is becoming stable and $D_{2}$ represents a phantom like fluid description while it is stable with accelerated nature of the universe.
\end{itemize}

\subsection{\label{sec:level II} Analysis of interacting R\'enyi HDE with interaction $Q=3H\rho_{d}$ :\protect}

Here, we study our said model with R\'enyi HDE considering a linear interaction $Q$ in the form of 
\begin{equation}
Q=3H\rho_{d}.
\end{equation}
By using (48), equations (23) and (24) transform into 
\begin{equation}
 \dot{\rho_{m}}=-3H\rho_{m}+3H\rho_{d}
 \end{equation}
 and 
 \begin{equation}
 \dot{\rho_{d}}=-6H\rho_{d}-3H\rho_{d}\omega_{d}
 \end{equation}
By incorporating the dimensionless variables defined in (25) and the equation (30), we obtain the following system of autonomous equation 

\begin{equation}
\begin{split}
\frac{dx}{dN}=3y+3xy\omega_{d} \\
\frac{dy}{dN}=-3y-3y(1-y)\omega_{d}
\end{split}
\end{equation}
Using (14) and (30), from (50), we obtain the equation of state parameter $\omega_{d}$ for R\'enyi HDE as 
\begin{equation}
\omega_{d}=\frac{\frac{2}{3}\frac{\dot{H}}{H^2+\pi\delta}}{1-2y}
\end{equation}
Here, we  again assume  $\lambda(x,y)=\frac{\dot{H}}{H^2+\pi\delta}$.
Here, we are assuming only one choice of 
\begin{equation}
\lambda(x,y)=\alpha x+\beta y
\end{equation}
as choice of an exponential function like previous case is not giving us any viable critical points.
By using (52) and  (53), our system of autonomous equations (51) transforms into 
\begin{equation}
\begin{split}
\frac{dx}{dN}=3y+\frac{2xy(\alpha x+\beta y)}{1-2y} \\
\frac{dy}{dN}=-3y+\frac{-2y(1-y)(\alpha x+\beta y)}{1-2y} 
\end{split}
\end{equation}

The critical points of this  autonomous system  are enlisted in Table-VIII.

\begin{table}[htbp]
    \centering
    \scriptsize
    \resizebox{\linewidth}{!}{
    \begin{tabular}{|c|c|}
        \hline
       \tiny{\textbf{Critical}} & \tiny{\textbf{Coordinates}}\\
        \tiny{\textbf{Points}} & \ \tiny{\textbf{$(x, y)$}} \\ \hline
        \tiny{$E$} &  \tiny{$(0, 0)$} \\ \hline
        
         \tiny{$E^{'}$} &  \tiny{$(0, 1)$} \\ \hline
         
         \tiny{$E_{1}$} &  \tiny{$(1-\frac{3+2\alpha-\beta+\sqrt{9+6\alpha+\beta^2}}{2(\alpha-\beta)}, \frac{3+2\alpha-\beta+\sqrt{9+6\alpha+\beta^2}}{2(\alpha-\beta)})$} \\ \hline
        \tiny{$E_{2}$} &  \tiny{$(1-\frac{3+2\alpha-\beta-\sqrt{9+6\alpha+\beta^2}}{2(\alpha-\beta)}, \frac{3+2\alpha-\beta-\sqrt{9+6\alpha+\beta^2}}{2(\alpha-\beta)})$} \\ \hline
     
    \end{tabular}
    }
    \caption{Set of critical points and their coordinates.}
\end{table}

Now, we study the characteristics of the system around the critical points and we will also study the evolution of the cosmological parameters.

\begin{itemize}
  \item  For critical point $E$ and $E^{'}$, we evaluate the eigen values from the linearized Jacobian matrix which is formed from the autonomous system (51). Both the eigen values corresponding to $E$  are $(-3,0)$ and eigen values corresponding to $E^{'}$ are $(-2\alpha-3,0)$. \\
  So, here both these critical points are non-hyperbolic in nature. Hence, we won't study them in this particular work.

\item It is very difficult to calculate the eigen values corresponding to  $E_{1}$ and $E_{2}$. For this we are having different choices of the values of parameters $\alpha,~\beta$ and corresponding to those values, we are representing respective points associated to $E_{1}$ and $E_{2}$ along with their eigen values and different cosmological parametric values in Table-IX and Table-X respectively.

\begin{table}[htbp]
    \centering
    \scriptsize
    \begin{adjustbox}{width=\linewidth}
    \begin{tabular}{|c|c|c|c|c|c|}
        \hline 
        \tiny \textbf{Choices of $(\alpha,\beta)$} & 
        \tiny \textbf{$E_1$} & 
        \tiny \textbf{Eigen Values} & 
        \tiny \textbf{$\Omega_{m}$} & 
        \tiny \textbf{$\Omega_{d}$} &  
        \tiny \textbf{$\omega_{d}$} \\ \hline 

          \tiny \textbf{$(15,1)$} & 
        \tiny \textbf{$(-\frac{1}{2},\frac{3}{2})$} & 
        \tiny \textbf{$(9,15)$} & 
        \tiny \textbf{$-\frac{1}{2}$} & 
        \tiny \textbf{$\frac{3}{2}$} &  
        \tiny \textbf{$2$} \\ \hline
        
         \tiny \textbf{$(6,-2)$} & 
        \tiny \textbf{$(-\frac{1}{2},\frac{3}{2})$} & 
        \tiny \textbf{$(\frac{21}{2},9)$} & 
        \tiny \textbf{$-\frac{1}{2}$} & 
        \tiny \textbf{$\frac{3}{2}$} &  
        \tiny \textbf{$2$} \\ \hline

        \tiny \textbf{$(-3,-5)$} & 
        \tiny \textbf{$(-\frac{1}{2},\frac{3}{2})$} & 
        \tiny \textbf{$(6,9)$} & 
        \tiny \textbf{$-\frac{1}{2}$} & 
        \tiny \textbf{$\frac{3}{2}$} &  
        \tiny \textbf{$2$} \\ \hline

          \tiny \textbf{$(4.5,0)$} & 
        \tiny \textbf{$(-1,2)$} & 
        \tiny \textbf{$(6,8)$} & 
        \tiny \textbf{$-1$} & 
        \tiny \textbf{$2$} &  
        \tiny \textbf{$1$} \\ \hline

    \end{tabular}
   \end{adjustbox}
    \caption{Eigen values and value of other cosmological parameters corresponding to $E_{1}$ for different choices of $\alpha$ and $\beta$}
\end{table}

\begin{table}[htbp]
    \centering
    \scriptsize
    \begin{adjustbox}{width=\linewidth}
    \begin{tabular}{|c|c|c|c|c|c|}
        \hline 
        \tiny \textbf{Choices of $(\alpha,\beta)$} & 
        \tiny \textbf{$E_2$} & 
        \tiny \textbf{Eigen Values} & 
        \tiny \textbf{$\Omega_{m}$} & 
        \tiny \textbf{$\Omega_{d}$} &  
        \tiny \textbf{$\omega_{d}$} \\ \hline 

          \tiny \textbf{$(15,1)$} & 
        \tiny \textbf{$(\frac{3}{14},\frac{11}{14})$} & 
        \tiny \textbf{$(-\frac{55}{2},-11)$} & 
        \tiny \textbf{$\frac{3}{14}$} & 
        \tiny \textbf{$\frac{11}{14}$} &  
        \tiny \textbf{$-4.66$} \\ \hline
        
         \tiny \textbf{$(6,-2)$} & 
        \tiny \textbf{$(\frac{3}{8},\frac{5}{8})$} & 
        \tiny \textbf{$(-35,-5)$} & 
        \tiny \textbf{$\frac{3}{8}$} & 
        \tiny \textbf{$\frac{5}{8}$} &  
        \tiny \textbf{$-2.66$} \\ \hline

        \tiny \textbf{$(-3,-5)$} & 
        \tiny \textbf{$(\frac{3}{2},-\frac{1}{2})$} & 
        \tiny \textbf{$(-2,1)$} & 
        \tiny \textbf{$\frac{3}{2}$} & 
        \tiny \textbf{$-\frac{1}{2}$} &  
        \tiny \textbf{$-0.66$} \\ \hline

          \tiny \textbf{$(4.5,0)$} & 
        \tiny \textbf{$(\frac{1}{3},\frac{2}{3})$} & 
        \tiny \textbf{$(-24,-6)$} & 
        \tiny \textbf{$\frac{1}{3}$} & 
        \tiny \textbf{$\frac{2}{3}$} &  
        \tiny \textbf{$-3$} \\ \hline

    \end{tabular}
   \end{adjustbox}
    \caption{Eigen values of  other cosmological parameters corresponding to $E_{2}$ for different choices of $\alpha$ and $\beta$}
\end{table}

\begin{figure}[htbp]
\centerline{\includegraphics[scale=0.6]{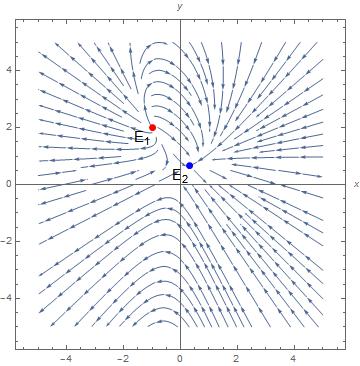}}
\caption{Phase plot corresponding to the point $E_{1}(-1,2)$ and  $E_{2}(\frac{1}{3},\frac{2}{3})$ for $\alpha=4.5, \beta=0$}
\end{figure}

  Fig-12 shows that  $E_{1}(-1,2)$ is a saddle node  while $E_{2}(\frac{1}{3},\frac{2}{3})$ is locally stable for $\alpha=4.5, \beta=0$.

\begin{figure}[htbp]
\centerline{\includegraphics[scale=0.6]{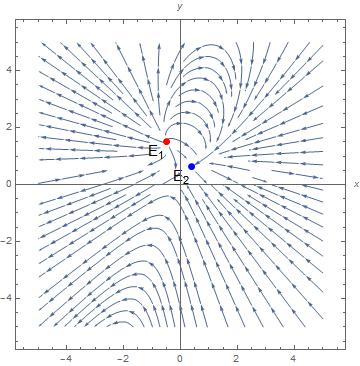}}
\caption{Phase plot corresponding to the point $E_{1}(-\frac{1}{2},\frac{3}{2})$ and  $E_{2}(\frac{3}{8},\frac{5}{8})$ for $\alpha=6, \beta=-2$}
\end{figure}
 Fig-13 shows that  $E_{1}(-\frac{1}{2},\frac{3}{2})$ is an unstable node  while $E_{2}(\frac{3}{8},\frac{5}{8})$ is locally stable for $\alpha=6, \beta=-2$.\\
  From here, we can comment that  $E_{1}$ is unstable while $E_{2}$ is  stable for any $\beta$ when $\alpha>0$.
\end{itemize}

\subsection{\label{sec:level II} Analysis of interacting R\'enyi HDE with interaction $Q=3H\rho_{m}$ :\protect}
Here, we consider a linear interaction $Q$ in the form of 
\begin{equation}
Q=3H\rho_{m}.
\end{equation}
By using (55), equations (23) and (24) transform into 
\begin{eqnarray}
 \dot{\rho_{m}}=-3H\rho_{m}+3H\rho_{m} \\ \nonumber
 \implies \dot{\rho_{m}}=0
 \end{eqnarray}
 and 
 \begin{equation}
 \dot{\rho_{d}}=-3H\rho_{d}-3H\rho_{d}\omega_{d}-3H\rho_{m}
 \end{equation}
Employing the dimensionless variables defined in (25) and the equations (30), (56) and (57), we obtain the following system of autonomous equation 

\begin{equation}
\begin{split}
\frac{dx}{dN}=3x+3xy\omega_{d} \\
\frac{dy}{dN}=-3x-3y(1-y)\omega_{d}
\end{split}
\end{equation}
Using (14) and (30), from (57), we obtain the equation of state parameter $\omega_{d}$ for R\'enyi HDE as 
\begin{equation}
\omega_{d}=\frac{1+\frac{2}{3}\frac{\dot{H}}{H^2+\pi\delta}-\frac{x}{y}}{1-2y}
\end{equation}
Similar to the previous occasions, we assume  $\lambda(x,y)=\frac{\dot{H}}{H^2+\pi\delta}$.
Here, we are assuming only one choice of 
\begin{equation}
\lambda(x,y)=\alpha x+\beta y
\end{equation}
as like previous interacting cases, choice of an exponential function is not giving us any viable critical points.
By using (59) and  (60), our system of autonomous equations (58) changes to
\begin{equation}
\begin{split}
\frac{dx}{dN}=3x+\frac{2xy(\alpha x+\beta y)+3xy-3x^2}{1-2y} \\
\frac{dy}{dN}=-3x+\frac{-2y(1-y)(\alpha x+\beta y)-3y(1-y)+3x(1-y)}{1-2y} 
\end{split}
\end{equation}

The critical points of this  autonomous system  are given in Table-XI.

\begin{table}[htbp]
    \centering
    \scriptsize
    \begin{tabular}{|c|c|}
        \hline
       \textbf{Critical} & \textbf{Coordinates}\\
        \textbf{Points} & \textbf{$(x, y)$}\\ \hline
        $F_1$ &  $(0, 0)$ \\ \hline
        
         $F_2$ &  $(0, 1)$ \\ \hline

         $F_3$ &  $(0, -\frac{3}{2\beta})$ \\ \hline
     
    \end{tabular}
    \caption{Set of critical points and their coordinates.}
\end{table}

Eigen values corresponding to all critical points are given in the Table-XII

\begin{table}[htbp]
    \centering
    \scriptsize
    \begin{tabular}{|c|c|}
        \hline
       \textbf{Critical} & \textbf{Eigen}\\
        \textbf{Points} & \textbf{values}\\ \hline
        $F_1$ &  $(-3, 3)$ \\ \hline
        
         $F_2$ &  $(-2\beta-3, -2\beta)$ \\ \hline

         $F_3$ &  $(3, \frac{9+6\beta}{6+2\beta})$ \\ \hline
     
    \end{tabular}
    \caption{Eigen values pertaining to all critical points of autonomous system (61).}
\end{table}

\begin{itemize}
\item $F_1$ always represents a saddle. In this case, we are unable to draw any conclusion regarding the acceleration of the universe as the equation of state parameter remains undetermined here.
\item $F_2$ becomes stable when $\beta>0$, represents a saddle when $-\frac{3}{2}<\beta<0$ and becomes unstable when $\beta<-\frac{3}{2}$. Stable $F_2$ always indicates a phantom like fluid description as the equation of state parameter, $\omega_d<-1$. In this case, the universe is completely dark energy dominated as the dark energy density corresponding to the R\'enyi HDE with the chosen interaction is, $\Omega_d=1$.  
\item $F_3$ is unstable when $\beta>-\frac{3}{2}$ and saddle when $\beta<-\frac{3}{2}$.
\end{itemize}

\begin{figure}[htbp]
\centerline{\includegraphics[scale=0.6]{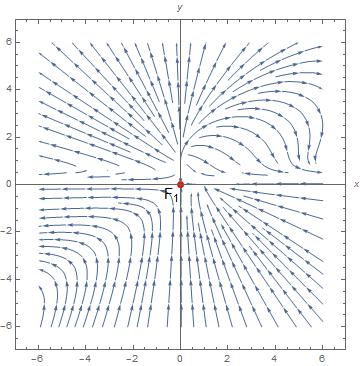}}
\caption{Phase plot corresponding to the point $F_{1}(0,0)$ for $\alpha=2, \beta=-3$}
\end{figure}

  Fig-14 shows that $F_{1}(0,0)$ represents a saddle.

\begin{figure}[htbp]
\centerline{\includegraphics[scale=0.9]{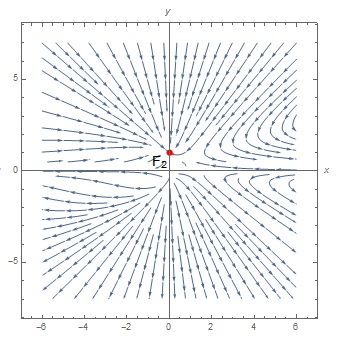}}
\caption{Phase plot corresponding to the point $F_{2}(0,1)$ for $\alpha=-1, \beta=4$}
\end{figure}

\begin{figure}[htbp]
\centerline{\includegraphics[scale=0.9]{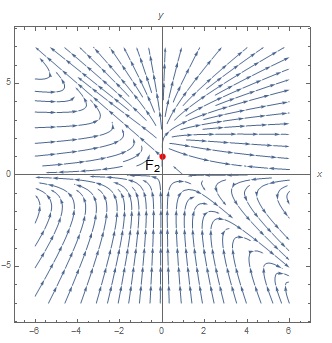}}
\caption{Phase plot corresponding to the point $F_{2}(0,1)$ for $\alpha=-1, \beta=-1$}
\end{figure}
Fig-15 shows that $F_{2}(0,1)$ is stable when $\beta$ is positive while Fig-16 shows that $F_{2}(0,1)$ is saddle or unstable for negative values of $\beta$.
\begin{figure}[htbp]
\centerline{\includegraphics[scale=0.6]{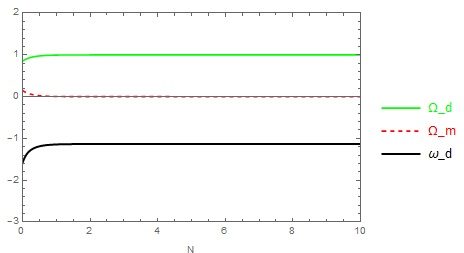}}
\caption{Evolution of cosmological parameters near $F_{2}(0,1)$ for $\alpha=-1, \beta=0.2$}
\end{figure}
Fig-17 shows that $F_2$ is indicating towards a dark energy dominated era and the equation of state parameter evolving towards -1 but completely denoting the phantom description of cosmic fluid.

\subsection{\label{sec:level II} Analysis of interacting R\'enyi HDE with non-linear interaction $Q=3H\frac{\rho_{d}}{\rho_{m}+\rho_{d}}$ :\protect}

Here, we study our said model with R\'enyi HDE considering a non linear interaction $Q$ in the form of 
\begin{equation}
Q=3H\frac{\rho_{d}}{\rho_{m}+\rho_{d}}
\end{equation}
By using (62), equations (23) and (24) are transformed into 
\begin{equation}
 \dot{\rho_{m}}=-3H\rho_{m}+3H\frac{\rho_{d}}{\rho_{m}+\rho_{d}}
 \end{equation}
 and 
 \begin{equation}
 \dot{\rho_{d}}=-3H\rho_{d}-3H\rho_{d}\omega_{d}-3H\frac{\rho_{d}}{\rho_{m}+\rho_{d}}
 \end{equation}
To frame an autonomous system of differential equations, we introduce new variables 
\begin{equation}
x=\rho_{m},~~~y=\rho_{d}
\end{equation}
Hence the energy density parameters will be in the following form
\begin{equation}
\Omega_{m}=\frac{x}{x+y},~~~\Omega_{d}=\frac{y}{x+y},~~~~3H^2=x+y
\end{equation}
By using (63), (64) and (65), we construct the following system of autonomous equations
\begin{equation}
\begin{split}
\frac{dx}{dN}=-3x+\frac{3y}{x+y} \\
\frac{dy}{dN}=-3y-\frac{3y}{x+y}-3y\omega_{d}
\end{split}
\end{equation}
Using R\'enyi HDE with Hubble's cut off i.e. equation (14), from (64), we derive the expression for the equation of state parameter $\omega_{d}$ as 
\begin{equation}
\omega_{d}=-\frac{4}{3}\frac{\dot{H}}{H^2}+\frac{2}{3}\frac{\dot{H}}{H^2+\pi\delta}-1-\frac{1}{\rho_{m}+\rho_{d}}
\end{equation}
 From (19), we find the expression for 
 \begin{equation}
    \frac{\dot{H}}{H^2}=-\frac{3}{2} -\frac{3}{2}\frac{y}{x+y}\omega_{d}
 \end{equation}
 Using (68) and (69), the system of autonomous equations in (67) further transforms into 
 \begin{equation}
\begin{split}
\frac{dx}{dN}=-3x+\frac{3y}{x+y} \\
\frac{dy}{dN}=-3y-\frac{3y}{x+y}-\frac{3y(x+y-1)}{x-y}-\frac{2\frac{\dot{H}}{H^2+\pi\delta}(x+y)y}{x-y}
\end{split}
\end{equation}
Equation of the state parameter takes the form
\begin{equation}
\omega_{d}=\frac{x+y-1}{x-y}+\frac{\frac{2}{3}(x+y)\frac{\dot{H}}{H^2+\pi\delta}}{x-y}  
\end{equation}

Here, we put  $\lambda(x,y)=\frac{\dot{H}}{H^2+\pi\delta}$
and assume only one choice of 
\begin{equation}
\lambda(x,y)=\alpha x+\beta y
\end{equation}
as choice of an exponential function is not giving us any viable critical points.
By using (71) and (72), our system of autonomous equations (70) changes into 

 \begin{equation}
\begin{split}
\frac{dx}{dN}=-3x+\frac{3y}{x+y} \\
\frac{dy}{dN}=-3y-\frac{3y}{x+y}-\frac{3y(x+y-1)}{x-y}-\frac{2(\alpha x+\beta y)(x+y)y}{x-y}
\end{split}
\end{equation}

The critical point corresponding to this autonomous system can be found as given in Table-XIII

\begin{table}[htbp]
    \centering
    \scriptsize
    \begin{tabular}{|c|c|}
        \hline
        \textbf{Critical Points} & \textbf{Coordinates $(x, y)$} \\ \hline
        $G$ & $\left(\frac{\alpha}{\alpha - \beta},\ -\frac{\alpha^2}{(\alpha - \beta)\beta}\right)$ \\ \hline
    \end{tabular}
    \caption{Set of critical points and their coordinates.}
\end{table}

While we are trying to find the eigen values pertaining to the critical point $G$ from the linearized matrix obtained from the system of autonomous equations in (73), it becomes a very complicated one.

For that we are representing a tabular form of eigen values in Table-XIV corresponding to the critical point $G$ for different choices of $\alpha$ and $\beta$.

\begin{table}[H]
    \centering
    \scriptsize
    \begin{adjustbox}{width=\linewidth}
    \begin{tabular}{|c|c|c|c|c|c|}
        \hline 
        \tiny \textbf{Choices of $(\alpha,\beta)$} & 
        \tiny \textbf{$G$} & 
        \tiny \textbf{Eigen Values} & \tiny \textbf{$\Omega_{m}$} & \tiny \textbf{$\Omega_{d}$} & \tiny \textbf{$\omega_{d}$}
         \\ \hline 

          \tiny \textbf{$(1,2)$} & 
        \tiny \textbf{$(-1,\frac{1}{2})$} & 
        \tiny \textbf{$(\frac{-4-\sqrt{83}i}{3},\frac{-4+\sqrt{83}i}{3})$} &  \tiny \textbf{$2$} & \tiny \textbf{$-1$} &\tiny \textbf{$1$}
         \\ \hline
        
         \tiny \textbf{$(-2,-1)$} & 
        \tiny \textbf{$(2,-4)$} & 
        \tiny \textbf{$(\frac{5-\sqrt{313}}{6},\frac{5+\sqrt{313}}{6})$} &  \tiny \textbf{$-1$} & \tiny \textbf{$2$} &\tiny \textbf{$-0.5$}
       \\ \hline

        \tiny \textbf{$(4,1)$} & 
        \tiny \textbf{$(\frac{4}{3},-\frac{16}{3})$} & 
        \tiny \textbf{$(\frac{-35-\sqrt{67}\sqrt{5}i}{10},\frac{-35+\sqrt{67}\sqrt{5}i}{10})$} &  \tiny \textbf{$-\frac{1}{3}$} & \tiny \textbf{$\frac{4}{3}$} &\tiny \textbf{$-0.75$}
         \\ \hline

       \tiny \textbf{$(-1,-2)$} & 
        \tiny \textbf{$(-1,\frac{1}{2})$} & 
        \tiny \textbf{$(\frac{-2-\sqrt{59}i}{3},\frac{-2+\sqrt{59}i}{3})$} &  \tiny \textbf{$2$} & \tiny \textbf{$-1$} &\tiny \textbf{$1$}
         \\ \hline

         \tiny \textbf{$(-6,-2)$} & 
        \tiny \textbf{$(\frac{3}{2},-\frac{9}{2})$} & 
        \tiny \textbf{$(\frac{33-3\sqrt{329}}{8},\frac{33+3\sqrt{329}}{8})$} &  \tiny \textbf{$-\frac{1}{2}$} & \tiny \textbf{$\frac{3}{2}$} &\tiny \textbf{$-0.66$}
       \\ \hline

    \end{tabular}
   \end{adjustbox}
    \caption{Eigen values corresponding to $G$ for different choices of $\alpha$ and $\beta$}
\end{table}

\begin{figure}[h!]
\centerline{\includegraphics[scale=0.6]{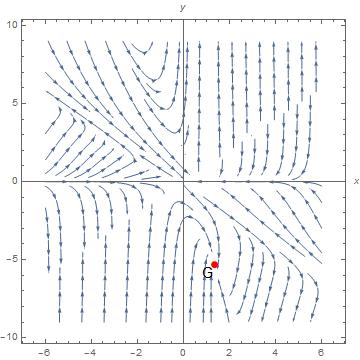}}
\caption{Phase plot corresponding to the point  $G(\frac{4}{3},-\frac{16}{3})$ for $\alpha=4, \beta=1$}
\end{figure}

\begin{figure}[H]
\centerline{\includegraphics[scale=0.6]{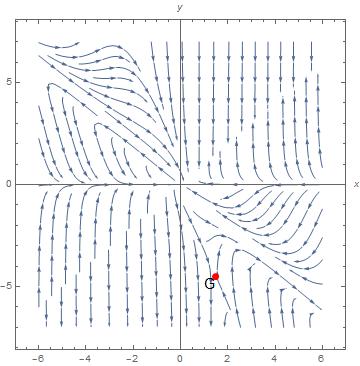}}
\caption{Phase plot corresponding to the point $G(\frac{3}{2},-\frac{9}{2})$  for $\alpha=-6, \beta=-2$}
\end{figure}

Fig-18 shows that  $G(\frac{4}{3},-\frac{16}{3})$ is a stable node while Fig-19 shows that  $G(\frac{3}{2},-\frac{9}{2})$ is an unstable node.

\section{Comparison with other R\'enyi Holographic dark energy models}
Maity et. al. \cite{maity2019} have found a stable solution for R\'enyi HDE model with Hubble's cut off. Their model is representing a quintessence behavior, whereas, we found that our R\'enyi HDE model with Hubble's cut off can represent both the quintessence and phantom behavior having stable nature. In their R\'enyi HDE model, Moradpour et. al. \cite{Moradpour2018} have also found that for some parametric choices, the cosmos may cross the phantom barrier and R\'enyi HDE model is stable in matter dominated era. We have also found for our interacting R\'enyi HDE model with the interaction $Q=3H(\rho_m+\rho_d)$ that the stable critical point $C$ indicates matter dominated era which matches with the previous observation. Dubey et. al. \cite{dubey2020} have shown that the RHDE model imitates the cosmological constant at far future. They have also shown that the equation of state parameter varies from a quintessence region to phantom region while in our model, either the equation of state parameter staying in phantom region or it is transiting from phantom to quintessence via cosmological constant. Jawad et. al. \cite{jawad2018} have also shown that the trajectories of equation of state parameter in their RHDE model show transition of the universe from phantom era (at early and present) towards quintessence era (latter epoch) by crossing phantom barrier which is similar to our observation. Vijaya Santhi et. al.\cite{VijayaSanthi2022} have found that the trajectories of the equation of state parameter for their models start from quintessence phase and turn towards phantom region by crossing phantom barrier.

\section{\label{sec:level V}Conclusion:\protect}

In this study, we consider a cosmological model of the universe filled with a perfect fluid within the framework of a spatially flat Friedmann–-Lemaître-–Robertson-–Walker (FLRW) metric. The universe is assumed to contain only dark matter and dark energy components, with the R\'enyi holographic dark energy serving as the sole source of dark energy. Both non-interacting and interacting scenarios between dark matter and dark energy have been examined.\\
To analyze the model's dynamical behavior, we employ the tools of dynamical systems theory. Accordingly, we construct an autonomous system to study the evolution and stability of the cosmological model around its critical points. In this formulation, we introduce a functional expression involving the Hubble parameter, given by $\frac{\dot{H}}{H^2+\pi \delta}$, as a key component of the system.\\
In the non-interacting case, we explore two functional forms-—linear and exponential of the dynamical variables, parameterized by two arbitrary constants $\alpha$ and $\beta$. These choices lead to rich and interesting dynamical behavior, offering insights into the evolution of the universe under the R\'enyi  holographic dark energy framework.\\
We find that, corresponding to the stable hyperbolic node $A$(as shown in Figures 1, 2, and 4), where a linear function of the dynamical variables is considered, the cosmological model exhibits interesting behavior depending on the choice of the parameter $\beta$. Specifically, for different values of $\beta$, the model is capable of describing both quintessence-like and phantom-like dark energy regimes, each associated with accelerated cosmic expansion.\\
Pertaining to the stable hyperbolic node $B$ (as shown in Fig-6 and 7) where an exponential function of the dynamical variables is considered, the cosmological model is becoming stable for any choice of $\beta$ but the fluid description is always indicating towards phantom era with accelerated expansion.\\
In both these linear and exponential functions our non-interacting R\'enyi HDE model indicates a dark energy dominated era (as shown in Fig-5 and Fig-8) as well as we can conclude that the parameter $d$ which is considered in the R\'enyi HDE energy density does not impact much in our model but the value of  $\delta$ is negative for some choices of parameters. The evolution of equation of state parameter (as shown in Fig-5) for the choice of linear function shows a shifting from phantom era to quintessence era.\\
 For, interacting R\'enyi HDE, we have assumed four interactions as  
 \begin{itemize}
  \item $Q=3H(\rho=\rho_{m}+\rho_{d})$.
  \item $Q=3H\rho_{d}$.
  \item $Q=3H\rho_{m}$.
  \item $Q=3H\frac{\rho_{d}}{\rho_{m}+\rho_{d}}$.
\end{itemize}

For the 1st interaction, we are having a stable node $C$ (as shown in Fig-9) which is stable for any choice of $\alpha>0$ but nothing can be told regarding the fluid description or the acceleration here. This node indicates a matter dominated universe. The node $D_1$ is unstable in most cases (as shown in Fig-10,11) for different choices of parameters $\alpha$ and $\beta$ where as $D_2$ (as shown in Fig-10,11) is stable for positive $\alpha$ and $\beta$ and indicates a phantom era with accelerated expansion.\\
 Corresponding to 2nd interaction, similarly hyperbolic node $E_1$ is  unstable in most cases and $E_2$ is stable for any $\beta$ when $\alpha>0$ (as shown in Fig-12,13). Stable $E_2$ indicates a phantom era with accelerated expansion.\\
In the 3rd linear interaction, we have considered that interaction is proportional to $\rho_{m}$. Here critical point $F_2$ seems interesting as it is becoming stable for positive values of the parameter $\beta$ (as shown in Fig-15) and it is indicating a phantom era with dark energy domination (as shown in Fig-17).\\ 
  Pertaining to 4th interaction which is non-linear in nature, node $G$ is stable for some choices of parameters $\alpha$ and $\beta$  and some stable node indicates towards an accelerated expansion with quintessence era.\\
  In summary, the non-interacting R\'enyi Holographic Dark Energy  model exhibits the capability to describe both quintessence and phantom eras, with a clear indication of dark energy domination in the late-time evolution of the universe. When interactions are introduced, the nature of the interaction function plays a significant role: our chosen linear interactions tend to favor a phantom-like behavior, while nonlinear interaction typically leads to a quintessence-like regime.

  Importantly, in nearly all scenarios explored—whether interacting or non-interacting—-the R\'enyi HDE model demonstrates a stable dynamical behavior characterized by a late-time accelerated expansion. This robustness highlights the potential of R\'enyi entropy-based dark energy as a viable candidate for explaining the current accelerated phase of the universe.

\section{Acknowledgement}
Authors are thankful to Prof. Subenoy Chakraborty for his comments and suggestion during a discussion.

\bibliography{Paper}
\end{document}